\documentclass{aa}
\usepackage{graphicx}
\usepackage{txfonts}
\usepackage{floatflt}
\usepackage{natbib}
\newcommand{\kms}{km\,s$^{-1}$}
\begin{document}
  \title{An analysis of a spectrum of V838 Monocerotis in October 2005}

  \author{R. Tylenda, T. Kami\'{n}ski, and M. Schmidt} 

  \offprints{R. Tylenda}

  \institute{\center Department for Astrophysics, N. Copernicus
            Astronomical Center, 		  Rabia\'{n}ska 8,
            87-100 Toru\'{n}, Poland\\
   \email{tylenda@ncac.torun.pl, tomkam@ncac.torun.pl, schmidt@ncac.torun.pl}}
 
 \date{Received; accepted}

\abstract{V838 Mon erupted at the beginning of 2002 becoming an extremely
 luminous star with $L=10^6~L_{\sun}$. Among various scenarios
 proposed to explain the nature of the outburst, the most promising is
 a stellar merger event.}
 {We attempt to obtain information about the structure and evolution of the
 object in the decline from the 2002 eruption.}
{The results of spectroscopic observations of the
object obtained in October~2005 with the Keck/HIRES instrument, presented in
detail in Paper~I, are analysed and discussed. Modelling of the observed
line profiles has been used to constrain the physical parameters of the system.}
{The kinematics of the atmosphere of V838~Mon is very complex. 
Our analysis of the molecular bands and the P-Cyg profiles of
atomic lines shows that the object loses matter with a
velocity of up to 215~\kms\ and a rate of $10^{-6}-10^{-5}$\,M$_{\sun}$\,yr$^{-1}$.
In the profiles of some atomic lines, we have, however, found evidence of
matter infall. Moreover, a narrow absorption component, which is particularly strong
in some P-Cyg profiles, may indicate that a jet-like outflow has also been formed.

We show that the observed emission in the [\ion{Fe}{II}] lines and an
eclipse-like event observed in November/December~2006 was probably caused by
interactions of the expanding matter, ejected by V838~Mon in 2002, with
radiation from the B3V companion. In particular, the observed profiles of the
[\ion{Fe}{II}] lines can be easily modelled in this scenario and allow us to
estimate parameters of the system, such as the position of the B3V companion
relative to V838~Mon and the line of sight, density in the outflowing
matter, and mass lost in the 2002 eruption. The observed appearance of strong
H$\alpha$ emission, just before and during the eclipse-like event, can be
interpreted as a result of the accretion of the outflowing
matter onto the B3V companion: the accreted matter, shocked above the stellar
surface, can be a source of extreme-UV and soft X-ray
radiation capable of ionizing and exciting H in the outflow.}{}


\titlerunning{V838 Mon in 2005}

\authorrunning{Tylenda et al.}
 
\maketitle


\section{Introduction \label{intro}}

The eruption of V838 Monocerotis was discovered at the beginning of January~2002.
As observed in the optical, the eruption lasted about three
months \citep{muna02,kimes02,crause03}. During the event, the object reached 
a luminosity of $\sim 10^6~L_\odot$ . After developing an A--F supergiant spectrum 
at the maximum in the beginning of February~2002, the object evolved to
lower effective temperatures and in April~2002 was practically unable to be
detected in the optical, remaining very bright however in the infrared. Optical
spectroscopy acquired at that time discovered a B3V companion of the erupted
object \citep{mdh02}. \cite{tyl05} analysed
the evolution of V838~Mon during outburst and early decline.

Different outburst mechanisms, including an unusual nova, a He-shell flash,
and a stellar merger, were
proposed to explain the eruption of V838~Mon. 
These mechanisms were critically discussed by \cite{tylsok06},
the authors conclude that the only mechanism that
could satisfactorily account for the observational data was a collision and
merger of a low-mass pre-main-sequence star with an $\sim 8\,M_\odot$
main-sequence star.

In \cite{pap_I} (hereinafter referred to as Paper~I), a high resolution spectrum 
of V838~Mon acquired with the Keck~I telescope in October 2005 was
presented. In the
present paper, we analyse and discuss the results obtained from this
spectrum.


\section{Evolution of V838~Mon during the decline after the 2002 eruption
\label{decline_sect}}

 A few months after the eruption, discovered at the beginning of January~2002,
V838~Mon entered a relatively calm decline phase. V838~Mon then resembled a
very cool oxygen-rich (C/O~$<1$) supergiant slowly declining in luminosity
\citep{evans03,tyl05,mun_asp}.
It dominated the observed spectrum at green, red and infrared
wavelengths. The blue part of the spectrum was dominated by the light of 
the B3V companion.

Late in 2004, narrow emission lines, belonging mostly to
[\ion{Fe}{II}], appeared mainly in the blue part of the
spectrum \citep{barsuk06}, and strengthened in time
\citep{mun07}. During November--December~2006, an eclipse-like event
of the B3V companion was observed
(\citealp{bond06,mun07}). At the epoch of the event,
the [\ion{Fe}{II}] emission lines reached their maximum strength 
and strong emission in Balmer lines appeared \citep{mun07}.

In October 2005, we obtained an optical spectrum of V838~Mon using the
Keck/HIRES instrument. Results of these observations were presented in
Paper~I. 
The star V838~Mon itself is seen as a very cool supergiant that
dominates the green and red parts of the spectrum. Numerous, often
very deep and complex, 
molecular absorption bands are the main spectral characteristics of this component.
All the bands are from oxides and include TiO, VO, AlO, ScO, and YO. The
excitation temperature derived from the bands ranges from $\sim2500$~K,
identified also as the photospheric temperature of V838~Mon, down to $\sim500$~K,
which presumably corresponds to outflowing matter at a few stellar radii.
The most positive (heliocentric) radial velocity derived from the bands is
$\sim58$~km\,s$^{-1}$, which we propose to be the radial velocity of
V838~Mon itself. Several atomic lines, mostly resonance, display P-Cyg
profiles. They provide evidence of an intense mass outflow with a typical velocity of
$\sim150$~km\,s$^{-1}$. The blue part of the spectrum is dominated by the
spectrum of the B3V companion. The values of $T_{\rm eff}$ and log~$g$
derived from the spectrum agree well with those implied by the spectral type. 
The star is
a rapid rotator ($V$~sin\,$i \simeq 250$~km\,s$^{-1}$) and its (heliocentric) 
radial velocity is $\sim40$~km\,s$^{-1}$. 
Numerous emission lines were identified,
mainly in the blue part of the spectrum. They correspond predominantly to
[\ion{Fe}{II}] and exhibit the same profile, which can be fitted well with a
Lorentzian profile. The lines are centred on a heliocentric radial velocity
of $\sim13$~km\,s$^{-1}$ and have a FWHM of $\sim80$~km\,s$^{-1}$.

\cite{mun07} propose that the appearance of the [\ion{Fe}{II}] emission lines
and the eclipse-like event observed in November/December~2006 
were unrelated, and that the eclipse-like event
was caused by an eclipse in a binary system. We argue that both
events were strongly related and
produced by matter ejected from V838~Mon during 
the 2002 outburst and reaching the vicinity of the B3V companion (see
also \citealp{barsuk06,bond06}).

In principle, one can consider that because of a certain mechanism (e.g.,
dissipation of mechanical energy), the outer parts of the present V838~Mon wind
become excited, producing the observed [\ion{Fe}{II}] emission lines. 
In this case, however, the radial
velocity of the lines would be close to that of V838~Mon. This is
certainlly not the case. The [\ion{Fe}{II}] lines have a radial velocity of 
$13.3 \pm 0.7$\,km\,s$^{-1}$ (Sect.~4.1.1 in Paper~I), while the radial
velocity of V838~Mon is at least $53$\,km\,s$^{-1}$ (see
Sect.~\ref{v_rad}). In this case, it
would also be difficult to explain the observed [\ion{Fe}{II}] profiles.

All the observed characteristics of the [\ion{Fe}{II}] emission lines 
can, however, be easily explained, if they are assumed to be produced by
the matter ejected
during the 2002 eruption that approaches the B3V companion. As shown in
Sect.~\ref{feII_profil}, the observed line profiles as well as their radial
velocities can then be well accounted for. The continuous strengthening
of the lines in 2005--2006 is also easy to understand:
larger and larger amounts of matter become excited as it approaches the
source of excitation. Excitation by radiation from the B3V
companion then explains why the emission line spectrum is dominated by
\ion{Fe}{II}. In the spectrum of a B3V star, there are enough photons capable
of ionizing species with an ionization potential of $\sim 7$~eV, e.g., Fe, Ni,
while there are very few photons capable of ionizing the most abundant 
elements such as H, He or CNO. Finally, the idea easy explains the
$\sim 70$~day eclipse-like event observed in November/December~2006 
as an occultation 
by a dense cloud of the matter ejected in 2002 and now
crossing the line of sight of the B3V companion. 
If this is the case, one can
expect that the 2006 eclipse should be followed by similar events when other
fragments of the 2002 ejecta cross the line of sight. 
Indeed, a few months later V838~Mon entered
another eclipse-like event.\footnote{See e.g., the website of V. Goranskij
({\tt http://jet.sao.ru/$\sim$goray/v838mon.htm}).}
We note that the latter would be, 
at least, difficult to explain as a phenomenon typical of an eclipsing binary system.

\section{The radial velocity of V838 Mon}\label{v_rad}

Several molecules observed in the spectrum of V838~Mon reveal
bands formed from high excitation-energy levels (see Sect.~5 in Paper~I).
They are usually indicative of a radial heliocentric velocity
of $58\pm5$\,\kms\ (see also Sect.~\ref{bands}). 
It is reasonable to assume that they arise very close to the
photosphere and have a velocity close to the stellar one. Earlier
determinations of the stellar velocity in \cite{kolev} and
\cite{kipp04} implied a value of 59--62\,\kms, i.e., close to our result.

However, the SiO 
maser emission observed in V838~Mon \citep{degu,clauss,degkam} is at a heliocentric
velocity of 71\,\kms. In the case of late-type stars, the
SiO masers usually have a reliable measure of their stellar
radial velocity. 

There are two possible explanations of this discrepancy.
If the SiO maser emission really originates at the stellar velocity then the
estimate made in Paper~I may indicate that we do not measure 
molecular bands deep enough that are
to reach hydrostatic levels of the V838~Mon atmosphere, i.e., even the
highest excitation bands are formed in already expanding layers. The other
possibility is that the optical spectroscopic studies provide a correct
estimate
of the stellar velocity but that the SiO maser is not at the stellar velocity.
We note that the radial velocity and width of the main component of
the SiO maser are very close to those of the CO rotational lines observed in
diffuse matter in the close vicinity of V838~Mon \citep{kam08}. This point
certainly deserves further investigation.

In the present paper, we usually adopt the radial velocity derived from the
SiO maser, i.e., 71\,\kms, as the radial velocity of V838~Mon. One has,
however, to keep in mind that the results from optical spectroscopy
infer a value 10--15\,\kms\ lower.

\section{An analysis of the [\ion{Fe}{II}] line profile  \label{feII_profil}}

As discussed in Sect.~\ref{decline_sect}, we assume that the [\ion{Fe}{II}]
emission lines are formed in the matter ejected by V838~Mon during its 2002
outburst, which is now approaching the B3V companion and ionized by the UV radiation
of the companion. In this section, using simplified modelling, 
we show that the observed profiles of the [\ion{Fe}{II}] lines can be easily
explained by this scenario.

\subsection{The model  \label{feII_model}}

\begin{figure}
\includegraphics[angle=-90,scale=0.4]{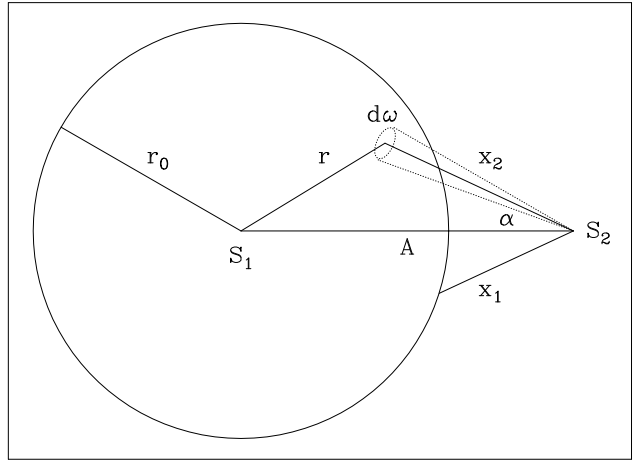}
  \caption{A sketch of the scenario considered in modelling the
  [\ion{Fe}{II}] line profile in Sect.~\ref{feII_profil}.}
\label{sketch}
\end{figure}	

We consider a system of two stars, $S_1$ and $S_2$, with a separation $A$ (see
Fig.~\ref{sketch}). 
We assume that the star $S_1$ is a source of a
steady, spherically symmetric wind, 
that forms an expanding envelope with a $r^{-2}$ density
distribution. We allow for the possibility that matter in the form of clumps or
filaments fills a $\xi$ part of the envelope volume, i.e., $\xi \le 1$.
The wind started a certain time ago, so the envelope has an
outer boundary, $r_0$. The wind has not yet reached star $S_2$, so $r_0 < A$.
Star $S_2$ is a source of ionizing radiation and emits 
$N_{\rm ph}$ ionizing photons isotropically per unit time and unit solid angle. The
radiation produces an ionized region in the wind whose boundaries
can be determined assuming ionization equillibrium.

We consider ionization along a given direction from star $S_2$ forming
an angle $\alpha$ with the axis between the stars. A distance $x_2$ from
star $S_2$, to which the wind matter is ionized, can be found from
\begin{equation}
  \label{ion_eq}
N_{\rm rec} = N_{\rm ph}\,{\rm d}\omega,
\end{equation}
where
\begin{equation}
  \label{rec_eq}
N_{\rm rec} = \int_{x_1}^{x_2} \xi\,x^2\,{\rm d}\omega\,\alpha_{\rm rec}\,
              N_{\rm i}^2\,{\rm d}x
\end{equation}
is a number of recombinations in a cone of solid angle d$\omega$ along 
the considered direction, $\alpha_{\rm rec}$ is a recombination coefficient,
while $N_{\rm i} = N_0\,(A/r)^2$ is an ion number density (electron
density is assumed to be equal to ion density).
In the following,
we assume that $A$ is a unit length, so all the $r$ and $x$ are expressed in
terms of $A$, and $x$ is then related to
$r$ and $\alpha$ via
\begin{equation}
  \label{x_eq}
  r^2 = 1 + x^2 - 2\,x\,\cos\alpha.
\end{equation}
In Eq. (\ref{rec_eq}), $x_1$ is a distance from
star $S_2$ to the wind outer boundary along the considered direction, given
by
\begin{equation}
  \label{x1_eq}
x_1 = \cos \alpha - \sqrt{\cos^2\alpha - (1-r_0^2)}.
\end{equation}
Equation (\ref{ion_eq}) can be rewritten as
\begin{equation}
  \label{ion_eq2}
C_0 \equiv \frac{N_{\rm ph}}{A^3\,\xi\,N_0^2\,\alpha_{\rm rec}} =
    \int_{x_1}^{x_2} \frac{x^2}{r^4}\,{\rm d}x,
\end{equation}
which, using Eq. (\ref{x_eq}), can be evaluated to give
\begin{equation}
  \label{ion_eq3}
C_0 = I(x_2) - I(x_1),
\end{equation}
where
\begin{equation}
  \label{i_eq}
I(x) \equiv \frac{x\,\cos 2\alpha - \cos\alpha}{2\,r^2\,\sin^2\alpha}
     + \frac{1}{2\,\sin^3\alpha}\,\arctan\frac{x-\cos\alpha}{\sin\alpha}.
\end{equation}
Equation (\ref{ion_eq3}) can be solved numerically to obtain $x_2$. Along the
separation axis, i.e., when $\alpha = 0$, $x_1 = 1 - r_0$, while Eq.~(\ref{i_eq}) 
reduces to
\begin{equation}
  \label{i0_eq}
I(x) = \frac{1}{1-x} - \frac{1}{(1-x)^2} + \frac{1}{3(1-x)^3}.
\end{equation}
Solving Eq. (\ref{ion_eq3}) for a grid of $\alpha$ values in the range $0 \le
\alpha \le \arcsin r_0$, allows us to
obtain the shape and position of the ionization front in the wind. 
Matter is ionized between
the wind outer boundary and the ionization front. We note that the star
separation axis is a symmetry axis of the ionized region.
There are two free parameters 
in the above problem, i.e., $r_0$ and $C_0$.

We assume that the ionized region is isothermal, so that the emission line coefficient 
varies as $N_i^2$, and that the intrinsic line profile is Gaussian
characterized by a thermal and/or turbulent velocity, $V_{\rm t}$. Integrating
the intrinsic line profile over the ionized region and taking into account
the kinematic properties of the ionized wind, a final emission-line profile can be
obtained. Apart from the above-mentioned parameters determining the
ionization front, the resultant line profile depends on kinematic parameters
of the wind and the stellar system, which are: the wind expansion velocity,
$V_{\rm wind}$; the velocity of star $S_1$ (source of the wind) relative 
to the observer, $V_{\rm s}$; and
the angle between the stars separation axis and the line of sight, 
$\alpha_{\rm s}$.

\subsection{Fitting the model to the observations  \label{feII_fit}}

Some of the parameters in the above problem can be estimated from
observations. As discussed in Sect.~\ref{decline_sect},
we assume that close to the time of the November/December~2006 eclipse, 
matter ejected during the 2002
eruption of V838~Mon reached the vicinity of the B3V companion. 
Our observations were completed
in October~2005, so we can estimate that $r_0 \simeq 0.75$. During the 2002 outburst 
of V838~Mon, expansion velocities observed 
reached $\sim 600$~km\,s$^{-1}$, 
although most of mass loss occurred at 150--400~km\,s$^{-1}$
\citep{muna02,crause03,kipp04,tyl05}. As discussed in \cite{tyl05}, the most
intense mass loss occurred in March~2002, which was observed as an expanding
photosphere of velocity $\sim 270$~\kms. 
We therefore assume 
$V_{\rm wind} \simeq 250~{\rm km\,s}^{-1}$ in the present calculations. 
We also assume that the heliocentric radial
velocity of V838~Mon is $V_{\rm s} = 71~{\rm km\,s}^{-1}$ (see
Sect.~\ref{v_rad}). Thus, there
remain 3 free parameters, i.e., $C_0$, $\alpha_{\rm s}$, and $V_{\rm t}$,
which can be obtained by fitting the model profile to the observed profile
of the [\ion{Fe}{II}] emission lines. 

\begin{figure}
 \centering
 \includegraphics[scale=0.45]{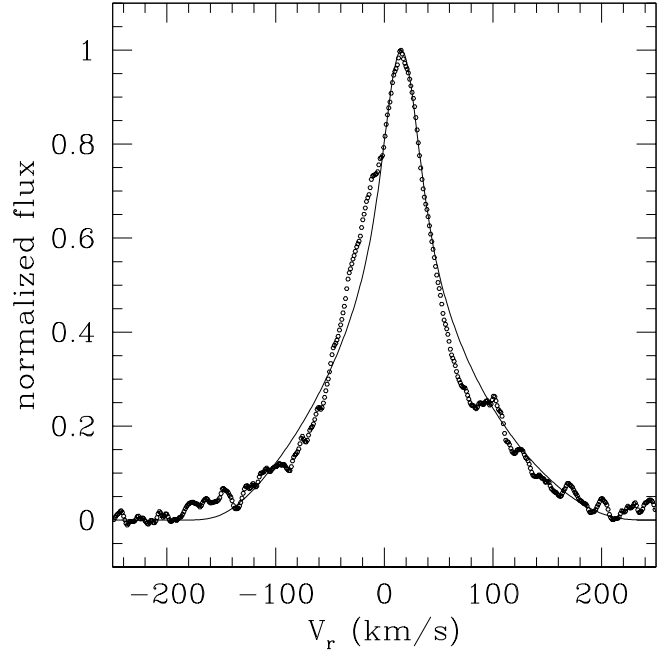}
 \caption{A model profile (full curve) compared to the observed profile of
[\ion{Fe}{II}]\,4287\,\AA\ (points). The model parameters are discussed in the text.} 
 \label{prof_fig}
\end{figure}

A fit of this kind is presented in Fig.~\ref{prof_fig}.
Points show the observed profile of the [\ion{Fe}{II}]\,4287\,\AA\ line. 
This is one of the
strongest emission lines of [\ion{Fe}{II}] in our spectrum. Its profile is free
of blending with other lines and affected little by noise. 
The local continuum was
subtracted and the flux was normalized to the peak value. (Note
that the profile of [\ion{Fe}{II}]\,4287\,\AA\, shown in Fig.~\ref{prof_fig}
is practically the same as the mean profile derived from 7 [\ion{Fe}{II}]
lines displayed in Fig.~4 of Paper~I.)
The model profile was obtained with the following values of the free
parameters: $C_0 = 0.35$, $\alpha_{\rm s} = 77\degr$, and 
$V_{\rm t} = 11.5~{\rm km\,s}^{-1}$. If $V_{\rm s} = 58~{\rm km\,s}^{-1}$ is
assumed (as discussed in Sect.~\ref{v_rad}),
the results of the profile fitting remain unchanged, except 
$\alpha_{\rm s}$, which increases to $80\degr$.

We note that there is little ambiguity in determining the values of 
the above free parameters, because each of them affects different characteristics of 
the model line profile. The value of
$C_0$ determines the size of the ionized region, so, for a
fixed $V_{\rm wind}$, it affects the wings of the line profile (the
higher the value of $C_0$, the more extended and stronger the wings).
If $V_{\rm wind}$ and $V_{\rm s}$ are fixed, $\alpha_{\rm s}$ defines the position of the
line peak. 
Finally, $V_{\rm t}$ determines the width of the line core.

\subsection{Discussion  \label{feII_disc}}

The results of the above modelling of the [\ion{Fe}{II}] profile
allows us to estimate the parameters of the matter
approaching the B3V companion. From the profile fitting, we have 
$C_0 = 0.35$. Assuming that the matter, expanding with velocity
$V_{\rm wind} \simeq 250~{\rm km\,s}^{-1}$,
reached the B3V companion in November or December of 2006, we derive a distance
between V838~Mon and the B3V companion of $A \simeq 3.7 \times 10^{15}$\,cm
($\sim 250$~AU).
Integrating the spectrum of a standard B3V star above the ionization
potential of iron, one obtains $N_{\rm ph} = 
2.4 \times 10^{46}\,{\rm s}^{-1}$. The recombination coefficient of
\ion{Fe}{I} is 
$\alpha_{\rm rec} \simeq 2.5 \times 10^{-12}\,{\rm cm}^3\,{\rm s}^{-1}$
(at an electron temperature of $10^3 - 10^4$\,K; see \citealp{nahar97}).
From Eq.~(\ref{ion_eq2}), one then finds that $N_0 \simeq 7 \times
10^5\,\xi^{-1/2}\,{\rm cm}^{-3}$, or $N_i \simeq 1.3 \times
10^6\,\xi^{-1/2}\,{\rm cm}^{-3}$ at the outer edge of the wind envelope,
i.e., at $r_0$. We note that at an electron density 
$\ga 10^7\,{\rm cm}^{-3}$, the [\ion{Fe}{II}] lines become collisionally
de-excited. 
Assuming spherical symmetry, a mass-loss rate during the
2002 outburst can be estimated to be
\begin{equation}
 \dot{M}_{\rm wind} = \frac{4\,\pi\,A^2\,V_{\rm wind}\,\xi\,N_0\,1.4\,m_{\rm H}}
                      {N_{\rm Fe}/N_{\rm H}}, 
\end{equation}
where $N_{\rm Fe}/N_{\rm H}$
is the relative-to-H number density of elements with an ionization potential 
similar to or lower than
that of Fe (i.e., Na, Mg, Al, Si, Ca, Ni), while $m_{\rm H}$ is the H atom
mass. Assuming solar abundances, $N_{\rm Fe}/N_{\rm H} \simeq 10^{-4}$ and
$\dot{M}_{\rm wind} \simeq 7 \times 10^{25}\,\xi^{1/2}\,{\rm g\,s}^{-1} 
\simeq 1\,\xi^{1/2}\,{\rm M}_{\odot}\,{\rm yr}^{-1}$. The most intense
period of mass
loss lasted a month (March) in 2002 \citep{tyl05}, so, according to the above
estimate, it produced a $\sim 0.1\,\xi^{1/2}\,{\rm M}_{\odot}$ shell. This
agrees with the total mass lost in the 2002 outburst, which was estimated to
have a mass
between $0.001 - 0.6\,{\rm M}_{\odot}$ in \cite{tyl05}.

\section{Mass outflow in V838 Mon} \label{outflow}

\cite{mun_asp} noted presence of continuous mass loss since the
outburst. They reported
that the outflow observed in \ion{K}{I} 7698~\AA\ has not much
changed since 2002. The nature of its mass loss
is however unclear. In this section, we analyse and
discuss results from the profiles of molecular bands and atomic lines
observed in October 2005 and presented in Paper~I.

\subsection{Outflow as seen in molecular bands} \label{bands}

\begin{figure}
 \centering
 \includegraphics[scale=0.45]{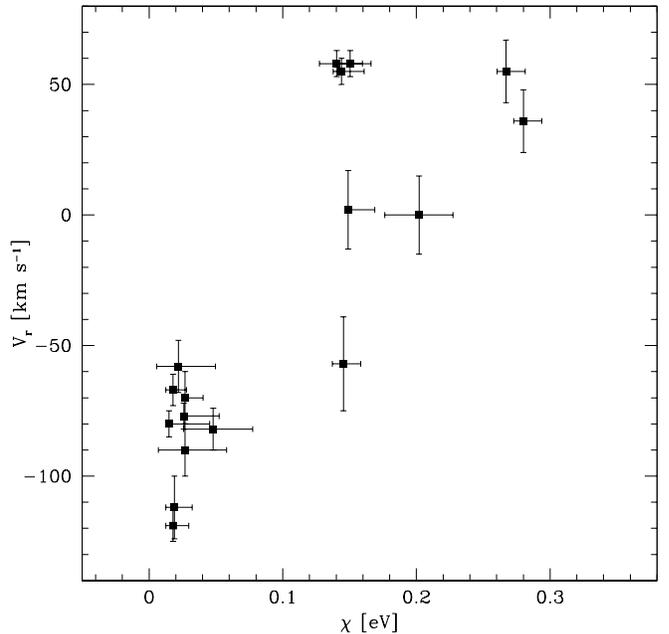}
 \caption{The heliocentric radial velocity of individual TiO bands plotted
 against the excitation potential of the lower level of the band. See text
 for more details.} 
 \label{bands_fig1}
\end{figure}

As discussed in Paper I, the spectrum of V838~Mon is dominated by numerous,
complex, and often very deep absorption lines related to molecular bands. A simple
model of a stellar photosphere plus an outflowing homogeneous layer was
used to identify and fit the observed structures of individual bands. In
this procedure, it was necessary to adopt various velocities for the
outflowing layer to reproduce different band structures. In general, we
found that bands originating in more excited levels correspond to
higher (more positive) radial velocities than low excitation bands. This
effect is shown, in a more quantitative way, in Fig.~\ref{bands_fig1}, which
plots the heliocentric radial velocity of a particular band versus
the excitation energy, $\chi$, of the level from which the band arises. Only results
for the TiO molecule, for which we identified the
largest number of bands, are shown. As the excitation potential plotted in
Fig.~\ref{bands_fig1}, we have taken the energy of the lower level of the
most blueshifted line in the bandhead. Vertical error bars illustrate
uncertainties in the velocity. Uncertainties in the energy have been taken
as a range of energies of lower levels of lines lying within
$\pm35$~\kms\ of the bandhead (this is the typical velocity broadening
adopted in modelling the bands in Paper~I).

As can be seen from Fig. \ref{bands_fig1}, there is a clear correlation
between the radial velocity of a particular band and the excitation
potential of the lower level of the band. This we interpret as evidence of
an outward accelerated mass outflow, in which the observed bands are
formed. The highest excitation bands can be formed in hotter regions, i.e.,
close to the photosphere, where the outflow velocity is low. Hence, we
postulated in Paper~I that the highest radial velocity observed in the
high excitation bands, i.e., $\sim +58$~\kms, corresponds to the radial
velocity of V838~Mon. The bands originating in levels close to the ground state 
can be easily formed in cold outermost layers outflowing with a large velocity. 

\begin{figure}
 \centering
 \includegraphics[scale=0.45]{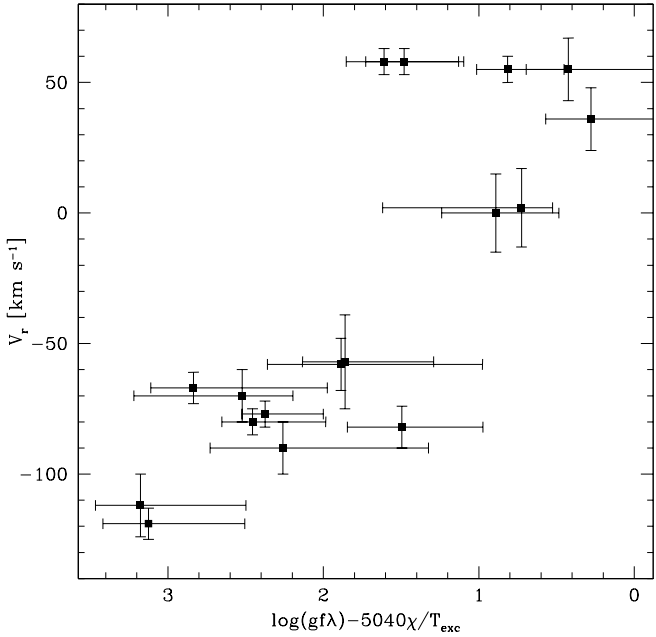}
 \caption{The heliocentric radial velocity of individual TiO bands plotted
 against log$(gf\lambda)-5040\chi/T_{\rm exc}$. See text
 for more details.} 
 \label{bands_fig2}
\end{figure}

Figure \ref{bands_fig2} shows the same data as in Fig.~\ref{bands_fig1} but in
a different, more astrophysical, way. Here, instead of the excitation
potential, we plot a parameter, log$(gf\lambda)-5040\chi/T_{\rm exc}$,
where $gf$ and $\lambda$ are the oscillator strength and wavelength of the band,
respectively. Within a constant factor, this parameter
is equivalent to the logarithm of
the absorption coefficient in a given band. As an
excitation temperature, $T_{\rm exc}$, we adopted 500~K, which is a
typical value in the outflow, as inferred in Paper~I (Sect.~5). The
horizontal error bars show uncertainties related to the spread in the excitation 
energy and oscillator strength for the lines present within 35~\kms\ of
the bandhead.

As Fig.~\ref{bands_fig1}, Fig.~\ref{bands_fig2} can be interpreted
as evidence of an increasing outflow velocity with the distance from the
photosphere. Bands with a small absorption coefficient require large column
densities to be visible in the observed spectrum. Thus, they must be
formed deep in the outflow. More opaque bands are effectively formed in
more distant layers and their profiles provide information about the kinematics
well above the photosphere.

From the span of the radial velocities observed in the TiO bands, shown
in Figs.~\ref{bands_fig1} and \ref{bands_fig2}, we can conclude
 that the wind in V838~Mon is accelerated at least
to a velocity of $\sim 180$~\kms.

\subsection{An analysis of the P-Cyg profiles}  \label{pcyg}

The presence of lines showing P-Cyg profiles in the spectrum of an
observed object is usually considered as
strong evidence of an ongoing mass outflow from the
object. In our spectrum of V838~Mon, we found about ten atomic
lines exhibiting these characteristics (see Sect.~4.2 in Paper~I). All of them
but one originate from resonant transitions. Below we analyse the
profiles of these lines in an attempt to investigate the physics and nature of
the outflow.

The terminal velocity of the outflow from V838~Mon can be obtained 
from the observed \ion{K}{I}\,$\lambda$7698 absorption profile (see Fig.~6 in Paper~I). 
If we take V$_s = 71$\,\kms\ to be the stellar
radial velocity (see discussion in Sect.~\ref{v_rad}), then the terminal
velocity of the outflow is V$_{\infty} = 215 \pm 5$~\kms. The observed
absorption components in other P-Cyg profiles do not reach this velocity,
most probably because of the
low column density of the atoms in the outer parts of the
outflow. Because of the high oscillator strength and the high
atomic abundance, the opacity in the \ion{K}{I}\,$\lambda$7698 line is at least two
orders of magnitude higher than for any other resonant line observed.
Consequently, this line can be observed even with relatively
low column densities expected at high velocities.
The value of terminal velocity derived from the
\ion{K}{I} absorption is consistent with, although slightly higher than, 
the outflow velocity estimated form the molecular bands of TiO analysed in
Sect.~\ref{bands}. It also agrees with the maximum expansion velocity derived by
\cite{geballe} from the CO bands observed half a year after our
observations.
 
To obtain a deeper insight into the nature of the outflow in
V838~Mon, we performed a radiative transfer modelling of the P-Cyg
profiles. We assume that the profiles are formed in a spherically
symmetric expanding envelope. Our modelling procedure is based on
the SEI method developed by \cite{sei}, and all details concerning the solution of
radiative transfer can be found therein\footnote{Aware of limitations
  of the SEI method we have also performed modelling in the more accurate
  comoving frame formalism \citep{cmf}. It confirms the results obtained with the
former method.}. As the velocity field of the
outflow, we adopt the standard $\beta$-law, i.e.,
\begin{equation}\label{betalaw}
w(x)=w_0+(1-w_0)\left(1-\frac{1}{x}\right)^{\beta},
\end{equation}
where $w=V/V_{\infty}$ is an outflow velocity normalized to the terminal velocity,
$x=r/{R_{\star}}$ is a dimensionless radial distance from the star with 
a photospheric radius $R_{\star}$, while $w_0$ is the normalized initial
velocity at the stellar photosphere (taken to be $w_0$=0.01 in our
calculations).  

As already pointed out in Paper~I (Sect.~4.2), the observed P-Cyg absorption
components have velocity structure. A majority of the absorption profiles
have narrow absorption components (NACs) that appear at different
velocities. The most striking example of NAC is seen in the \ion{Rb}{I}
line at V$_h$\,=\,--82~\kms (see Fig.~6 in Paper~I).
Some lines, e.g., those of \ion{Cr}{I} and
\ion{Ba}{I}, seem to exhibit many NACs at once, which dominate the
appearance of the profiles. On the other hand,
there are also absorption profiles, e.g., those of \ion{Ca}{I} $\lambda$6572 and
\ion{K}{I} $\lambda$7698, which are very smooth and do not seem to be affected by
discrete components. The question is whether the narrow components 
seen in our profiles are superimposed on a broad P-Cyg
absorption profile formed in a bulk flow, or whether the entire
absorption profile consists of several discrete and strong absorption
components that are blended and together form a more or less continuous profile.
       
As a first attempt, we assumed a distribution of
optical thickness in the form    
\begin{equation}\label{tau1}
\tau(w)={\mathcal T}_{\rm tot}\,(w/w_1)^{\alpha_1} \left(1-
        (w/w_1)^{1/\beta}\right)^{\alpha_2},
\end{equation}
where ${\mathcal T}_{\rm tot}=\int_{w_0}^{w_1}\tau(w)\,dw$ is the
integrated optical thickness and $w_1$ is the normalized velocity at
which opacity vanishes (usually taken to be 1). 
The values of ${\mathcal
  T}_{\rm tot}$, $w_1$, $\alpha_1$, and $\alpha_2$ were treated as
free parameters in our modelling procedure. The radial distribution
given by Eq.~(\ref{tau1}) is known to reproduce  ultraviolet (UV)
resonant lines formed in winds of hot stars \citep{tauref}. Our
attempts to reproduce the smooth profiles of
\ion{Ca}{I}, \ion{K}{I}, and the overall profile of \ion{Rb}{I}
(neglecting the NAC), showed that it was impossible to 
reproduce satisfactorily any of these lines with the opacity distribution in the
form of Eq.~(\ref{tau1}). The best set of the parameters found within this
approach resulted in profiles that were too narrow with respect to the
observed ones. 

We, therefore, modified the optical depth distribution by adding multiple 
Gaussian components, $\tau_i(w)$, to
Eq.~(\ref{tau1}) parameterized by
central velocities, widths, and weights according to which they are
added. We note that, for the sake of convenience, 
the Gaussian components are parameterized in 
the velocity domain, so the corresponding distributions as
functions of radial distance, $\tau_i(x)$, are in general not
Gaussian (the profile is modified by the adopted velocity field).  

\begin{figure}
\centering
\includegraphics[angle=-90,scale=0.7]{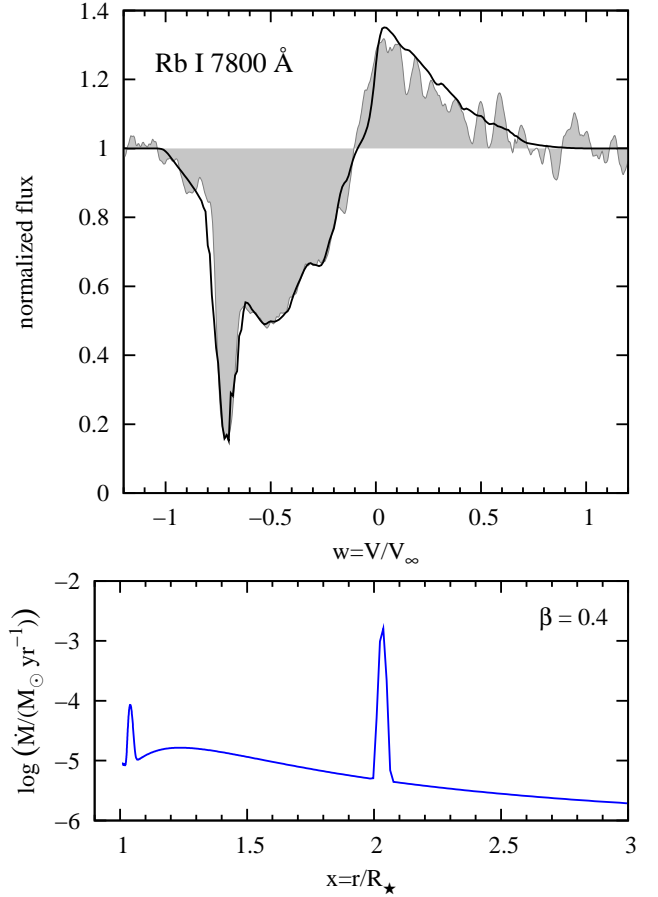}
  \caption{{\it Top:} The P-Cygni profile of \ion{Rb}{I} $\lambda$7800. 
The observed profile is drawn with a filled grey curve while the result of
the SEI
modelling is shown with a black line. {\it Bottom:} The mass loss rate 
corresponding to the fitted profile. The modelling was performed with
$\beta=0.4$ and $V_{\infty}=215$~km~s$^{-1}$.}
\label{rb1}
\end{figure}

In this multi-component approach, we were able to successfully reproduce 
the observed profiles.
Results for the \ion{Rb}{I} $\lambda$7800 line are shown in Fig.~\ref{rb1}.
At least
three discrete components (``shells'') of different widths are needed
to obtain a profile that reproduces the observations well. These
components, which are clearly present in all the profiles, are located at
outflow velocities of about 50, 110, and 150~\kms\ (with respect to 
V$_s$\,=\,71\,\kms). 

We attempted to model all the
prominent P-Cyg lines independently and, while the central velocities and 
widths of different components were found to be consistently
determined for different lines, the relative contributions of the
components differ considerably from line to line.
This implies that the relative ion opacities differ considerably between
consecutive ``shells''. This can be understand either as changes
in excitation conditions or different atomic abundances (chemical
composition) in the discrete outflow components. Although the latter
seems to be less likely, the changing atomic 
fractions due to interchanges of atoms with molecules and/or dust grains
may play an important role. An extreme case of this effect is 
observed in the
resonant absorption lines of \ion{Ti}{I}. Apart from the main component seen
close to the stellar velocity (but see Sect.~\ref{infall}),
these lines appear only
in one of the outflowing ``shells'' (see Fig.~6 in Paper~I). 

Our modelling allows us to place general constraints on
the velocity field in the outflow. The adopted value of
$\beta$ has a strong influence on the relative
strengths of absorption and emission components in a P-Cyg
profile (for a given radial distribution of optical thickness). 
We found that the outflow in V838~Mon can be characterized well by
$\beta=$0.35$\pm$0.05. This result was obtained assuming
that collisional deexcitation and photospheric contamination to the
P-Cyg profile can be neglected. An influence of collisional de-excitation
on the emission component cannot be properly recognized because of unknown
electron densities in the outflow. We believe that it is
small, at least for lines with high
transition probabilities (e.g., the lines of \ion{Rb}{I},
\ion{K}{I}, and \ion{Ba}{I} with an Einstein coefficient A$_{ul}$ of the
order of 10$^7$--10$^8$\,s$^{-1}$).
We did not find any signs of
photospheric contamination in the observed spectrum and we therefore 
neglected this effect in our modelling.      

The value of $\beta$ derived above implies that
the acceleration of material is very
rapid and at the distance of 4$R_{\star}$, the wind has a velocity of
0.99\,$V_{\infty}$. This steep velocity gradient requires very high
densities of material to be able to reproduce strong absorption features at low
velocities. Therefore,
to reproduce the absorption component
at the outflow velocity of 50\,\kms\ (which
corresponds to r\,=\,1.03\,$R_{\star}$ for $\beta=0.4$), it was
necessary to include an inner narrow shell with
a very high optical thickness. For several lines, e.g., the
\ion{K}{I} line, it was even impossible to reproduce the central parts
of the P-Cyg profile without
incorporating unreasonably high values of optical thickness. This made
us conclude that in the velocity field, generally well
described by the $\beta$-law, a sort of a plateau (or plateaus) may
exist, i.e., a region of nearly constant velocity extended in the
radial direction. 
These velocity plateaus, or non-monotonicities,
were proposed to explain the winds of hot stars \citep{nacs1,nacs2}.
The possible existence of
similar features in the outflow of V838~Mon is a tentative conclusion,
which should be examined by time-dependent observations of the P-Cyg
profiles in the spectrum of V838~Mon. 
The derived radial distributions of optical thickness for different ions
can be used to provide a rough estimate of the mass-loss rate in
V838~Mon. For a given stellar radius, $R_{\star}$, 
the mass-loss rate can be estimated from \citep{groen}
\begin{equation}\label{mdot}
\dot{M}(x)=\tau(x)\,w(x)\,x^2\,\frac{dw(x)}{dx}\,V_{\infty}^2\,R_{\star}
     \,\frac{4\,m_{\rm e}\,c}{e^2}\,\frac{\mu\,m_{\rm H}}{A_E\,f_{lu}\,\lambda_0},
\end{equation}
where $\mu$ is the mean molecular weight, $m_{\rm H}$ is the hydrogen
mass, $A_E$ is the abundance of 
an element with respect to hydrogen, $f_{lu}$ is the oscillator strength,
$\lambda_0$ is the laboratory wavelength of the line, and the symbols
$m_{e}$, $e$, and $c$ have their usual meanings. In Eq.~(\ref{mdot}), it is
assumed that all the atoms of a given element are in the 
ground state and that there is no ionization nor any other form of ion
depletion. Taking as the stellar radius $R_{\star}$\,=\,10$^3$\,R$_{\sun}$
(see Sect.5 in Paper~I),
adopting $\beta=0.4$, and assuming solar abundances, the mass-loss rate
derived from the lines of \ion{Rb}{I} $\lambda$7800 is on average of
$10^{-6} - 10^{-5}$\,M$_{\sun}$\,yr$^{-1}$,
with individual mass-loss events reaching 
$10^{-3}$\,M$_{\sun}$\,yr$^{-1}$ (see Fig.~\ref{rb1}).

\subsection{Signatures of an infall}\label{infall}
The redward broad absorption components seen in the \ion{Ti}{I} lines (see
Sect.~4.3 in Paper~I) can be interpreted as a signature of an infall in
V838~Mon. Similar absorption
components also contaminate the profiles
of the 5060 and 5110\,\AA\ lines of \ion{Fe}{I}.

These absorption features in the \ion{Ti}{I} lines have wings 
extending out to about $V_h$\,=\,120\,\kms. 
Thus, regardless of the velocity of the star in the
range 58--71\,\kms\ (see Sect.~\ref{v_rad}), 
the red wings of these features are evidently
redshifted with respect to the photosphere. The maximum infall velocity
inferred from these wings is 50--60\,\kms, which agrees very well with the
free fall velocity of an $8 M_\odot$ star \citep{tss05} 
at a photospheric radius of $\sim 900 R_\odot$ (Sect.~5 in Paper~I).

Signatures of an infall were previously observed in V838~Mon.
\cite{rush05} observed inverse P-Cyg profiles of \ion{Ti}{I} lines in
their spectra obtained in the near infrared in December~2003. From the separation
of the absorption and emission components, an infall velocity of
$\sim 20$~\kms\ can be deduced. These authors also argue that the infalling
matter, compressed, and heated when colliding with the atmospheric material,
provide conditions necessary to excite the SiO lines that they observe.
Observations of the ro-vibrational bands of CO by \cite{geballe} in April~2006
revealed a photospheric component redshifted by 15~\kms\ with respect to the velocity of
the SiO maser. This suggests, according to the authors, the presence of gas
infalling onto the star or contraction of the stellar photosphere.

\subsection{Discussion  \label{wind_disc}}

As mentioned above, the mass outflow in V838~Mon has been observed
continously
since the 2002 eruption. Our analysis confirms this statement and 
provides new results. The question that 
arises is what drives this outflow. It consists of cold neutral matter that
is rich in molecules, and possibly also dust, which is
similar to the composition of the winds of cool stars, e.g., objects in the asymptotic giant
branch. Velocity fields of these cool winds are usually well described by the
$\beta$-law with $\beta$\,$\simeq$\,0.5, which is close to the value
derived above for V838~Mon. These cool winds are believed to be
dust-driven. They usually do not reach the high terminal
velocities found for V838~Mon. A possible reason is that
dust is destroyed in collisions, when the outflow is faster than about
30\,\kms\ \citep{olo}. From this reason, it seems that the outflow in
V838~Mon cannot be driven by 
dust only, since we observe material being accelerated at much higher
velocities than 30\,\kms. This driving mechanism can, however, play a
role at the base of the outflow, where velocities are low enough. 

One can consider a mechanism similar to that driving winds in hot
stars, i.e., by radiation pressure absorbed in atomic resonant lines.
However, the small number and relative weakness of lines of this type in our spectrum
(they are typically 10 times narrower than those in the OB stars) does not
support the idea that these lines could be effective in driving mass loss from
V838~Mon. More promising is the possibility that the outflow
is driven by radiation pressure absorbed in molecular bands in
the optical and infrared. The object is very bright in these wavelength
ranges and the observed number and
strength of molecular bands, as seen in our spectrum (some of them
absorb practically all the radiation available in their wavelength ranges --
see Paper~I), clearly shows that a
significant part of the momentum carried out in radiation is indeed absorbed in
the outflowing matter. A process of driving winds
by radiation pressure absorbed in molecular bands was proposed 
as a mechanism accelerating
cold circumstellar envelopes \citep[see][and references therein]{jorg}. 

The analysis of the P-Cyg profiles reported in Sect.~\ref{pcyg} was
completed assuming spherical symmetry and a monotonicly increasing outflow
velocity described by Eq.~(\ref{betalaw}). Within this approach, the general
shape of the observed profiles can be satisfactorily reproduced. This
shows that the general pattern of the outflow is more or less spherically
symmetric. The profiles however show, at least some of them, narrow structures,
which we called NACs. The structure at $V_{\rm h} = -82$~\kms\ is
particularly striking. In the spherically symmetric, approach the only way to
explain the NACs is to assume that the mass-loss rate
varies with time and produces dense shells expanding with a thinner wind. As
can be seen in Fig.~\ref{rb1}, to explain the strong NAC at
$V_{\rm h} = -82$~\kms\ it is necessary to postulate a short lived
enhancement in the mass loss rate by two orders of magnitude. However, if
the above assumptions are relaxed, the NACs can be explained in
other ways. One possibility is that the NACs are produced by mass loss being more intense in
certain, more or less discrete, directions. For instance, the NAC at
$V_{\rm h} = -82$~\kms\  would then be understood as a stream-like or jet-like outflow
with a projected velocity of $\sim 153$~\kms\ 
(adopting the radial velocity of V838~Mon of 71~\kms; see Sect.~\ref{v_rad}).
Jets usually have counter-jets. The observed lack of any counter-jet in our
spectrum as well as the position of the NAC close to the terminal velocity
of the P-Cyg profiles suggests that the jet, if responsible for producing the NAC at 
$-82$~\kms, is not far from the line of sight.

Another possibility is that the NAC at $V_{\rm h} = -82$~\kms\ has nothing
to do with the presently ongoing mass outflow, but is produced in the matter
lost during the 2002 eruption that is now in front of the object. 
This interpretation would imply the existence of
a dense shell at a distance of $\sim 1.7\,\times\,10^{15}$~cm 
($\sim 2.5\,\times\,10^4\,\rm{R}_\odot$) from the object expanding with a
velocity of $\sim 153$~\kms. However, during the 2002
eruption the matter was ejected with velocities ranging
from $\sim 100$ to $\sim 600$~\kms\ \citep{muna02,crause03,kipp04}. Thus, the
matter, or at least part of it, would have to have been condensed into a shell
at later epochs. This can occur if matter ejected later catches up with slower
matter ejected earlier. In the case of V838~Mon, this mechanism does not seem
to have worked, since the outflow velocity observed during the
2002 eruption tended to decrease with time \citep{muna02,crause03}. 
A NAC at 
$V_{\rm h} = -90$~\kms\ is seen in our spectrum in the \ion{Ti}{I}~5147
and 5152~\AA\ lines, which do not display P-Cyg profiles 
(see panel~c in Fig.~6 in Paper~I). The close radial velocity and 
line width suggest that this component is produced in the
same enviroment as the NAC in the P-Cyg profiles.
If this is the case, our interpretation presented above
would have to be excluded. The \ion{Ti}{I} lines arise from the levels lying
2.3--2.4~eV above the ground state. Thus, they must arise in a rather warm
matter (of temperature comparable to that in the photosphere), while the matter lost in 2002 
and now at a distance of $\sim 30$ photospheric radii is expected to be cold
($\sim 500$~K, if estimated from pure geometric dilution of the radiative energy
density). In conclusion, we
consider that the interpretation that the NAC at $V_{\rm h} = -82$~\kms\ is
related to matter lost in 2002 is less
probable than the other two, i.e., varying mass-loss
rate in the present wind or jet-like ouflow.

Spectroscopic data allowing analyses of the time-dependent behaviour of the P-Cyg
profiles would help to discriminate between interpretations involving 
varying mass-loss rate, jet-like outflow, or distant matter lost in 2002. In
the first case, the NACs are expected to migrate in the profile with time,
while in the two other cases the NACs should be rather stable features.
It is worth noting that \cite{geballe} in their high-resolution spectra
obtained in the $K$ band in April~2006, i.e., half a year after our
spectroscopy, indentified several velocity
components in the CO absorption line profiles. Apart from the photospheric component
(mentioned in Sect~\ref{infall}) the authors found components at expansion
velocities of 15, 85, and 150~\kms. The last feature is practically at the
same velocity as our NAC at $V_{\rm h} = -82$~\kms\ suggesting that this is
a persistent component. However, the feature in the CO lines is much wider
than the NAC, so the conclusion is unclear. The two other components in
the CO lines do not match other NACs in our P-Cyg profiles, which are seen at
the expansion velocities of 50 and 110~\kms\ (see Sect.~\ref{pcyg}). This
suggests that these NACs are transient and produced by varying mass-loss
rates.

\section{Summary and discussion \label{sumdisc}}

Our analysis of both the positions of the molecular bands (Sect.~\ref{bands}) 
and the P-Cyg profiles (Sect.~\ref{pcyg}) have shown that V838~Mon loses
matter with a terminal velocity of $\sim 215$~\kms\ at a rate of
$10^{-6} - 10^{-5}$\,M$_{\sun}$\,yr$^{-1}$. Thus, the object loses of the
order of 10~$L_\odot$ in the form of the kinetic energy of the wind. This is
small compared to the radiation luminosity of the object, 
which is $\sim 2.4 \times 10^4~L_\odot$ (see Sect.~5 in Paper~I). 
However, the momentum carried out in
the wind is comparable to that in the radiation. Thus, the wind in V838~Mon
is likely to be radiation driven, probably due to absorption in the
molecular bands, as discussed in Sect.~\ref{wind_disc}.

The kinematic pattern of the matter in the
atmosphere of V838~Mon is very complex. Both expansion and wind outflow
dominate. Most of the outflow is more or less spherically symmetric
as suggested by our modelling of the observed P-Cyg profiles 
(Sect.~\ref{pcyg}). However, as discussed in Sect.~\ref{wind_disc}, 
it is quite possible that we also observe
a jet-like outflow. As shown in Sect.~\ref{infall}, 
certain regions, perhaps initially
accelerated but not sufficiently to allow them to leave the object, 
also fall back to the photosphere.
We note that simultaneous signatures of infall and outflow
have been observed in the winds of hot stars \cite[e.g.,][]{tsco} and often
in protostars. A possible scenario for creating coexisting mass outflow and
infall in asymptotic giant branch stars was discussed in \cite{sok08} 
and may also work in the remnant of the V838~Mon outburst.

As discussed in \cite{tyl05}, the evolution of V838~Mon 
in the post-outburst state is probably dominated by gravitational
contraction, i.e., is closely related to that of protostars.
If, as argued in \cite{tylsok06}, the 2002 outburst of V838~Mon was produced
by a stellar merger, then a significant amount of angular momentum should
have been stored in the remnant of the event \cite[see also][]{soktyl07}. 
Gravitational contraction may then result in the flattening of the envelope, eventually
leading to the formation of an accretion disc (as in protostellar objects).
The possibility discussed in Sect.~\ref{wind_disc}, namely  
that we observe a jet in the spectrum, if confirmed,
may imply that a disc has already formed in V838~Mon.

As shown in Sect. \ref{feII_profil}, the observed profile of the
[\ion{Fe}{II}] lines can be easily understood as having been produced by matter
lost by V838~Mon in the 2002 eruption and now excited by the radiation of 
the B3V companion. As the matter approaches the companion, a larger and larger
proportion of the flux from the star can interact with the matter causing an
increase in the [\ion{Fe}{II}] emission, as observed.
Indeed, the flux in the 20 strongest [\ion{Fe}{II}] lines in our spectrum
obtained in October~2005
increased by a factor of $\sim 3.5$ in the measurements completed by \cite{mun07}
in December~2006. At the moment of our observations,
the outer boundary of the expanding envelope was presumably at a distance of 
$\sim 0.75$ times the separation between V838~Mon and its B3V companion.
Thus, only $\sim 17$\% of the radiation flux of the
B3V companion should then interact with the matter. At the moment of the
eclipse-like event in November/December~2006, the matter presumably reached
the closest vicinity of the companion and the portion of the stellar
radiation capable of exciting the matter increased to $50-100$\%.

Apart from the [\ion{Fe}{II}] emission lines, \cite{mun07} also
observed strong Balmer lines in emission. In particular, they discussed in detail 
the profile of H$\alpha$ observed in December~2006 -- April~2007. No H$\alpha$
was seen in our spectrum in October~2005. However, in the absorption cores of
higher members of the Balmer series (in the spectrum of the B3V
companion), we have detected weak emission components (see Sect.~4.1.2 in
Paper~I). In particular, the emission component in the H$\beta$ line
was estimated to have an observed flux of $\sim 3 \times
10^{-16}$~erg\,cm$^{-2}$\,s$^{-1}$. We note that
the profile of the H$\beta$ emission component
(see Fig.~5 in Paper~I) was remarkably similar to that of the
H$\alpha$ line observed by \cite{mun07} in December~2006. Moreover, the
parameters of the emission component in H$\beta$ in our spectrum and of the
H$\alpha$ emission component in \cite{mun07} (radial velocity and FWHM) are
close to those of the [\ion{Fe}{II}] lines derived in Paper~I. This
implies that all these lines probably originate in the same region, i.e.,
in the part
of the V838~Mon envelope (ejected in 2002) approaching the B3V companion.
The point is, however, that while in the spectrum of a B3V star there are
enough photons capable of ionizing iron and exciting the [\ion{Fe}{II}] lines,
there are very few photons, capable of ionizing hydrogen.
Between October~2005 and December~2006, the
flux in the H$\beta$ emission also increased by a factor of $\sim 7$, while that in
H$\alpha$ increased by at least two orders of magnitude. This large increase 
cannot be explained by a purely geometric effect, as discussed above in the case 
of the [\ion{Fe}{II}] lines. Therefore, an additional source of excitation of
the Balmer lines is necessary, a source, whose intensity significantly increased
between October~2006 and December~2007. We suggest that this is provided by
matter accretion from the outflowing V838~Mon envelope on the B3V companion. 
We now complete simplified estimates for this scenario.

We assume that matter flows with a velocity, $V_{\rm wind}$, in
the vicinity of a star of mass, $M_{\rm B3V}$.
An accretion radius, $r_{\rm acc}$, can be estimated from the standard formula 
\citep[see e.g.,][]{fkr}
\begin{equation}
\label{racc}
  r_{\rm acc} \simeq \frac{2\,G\,M_{\rm B3V}}{V_{\rm wind}^2} 
              \simeq 3 \times 10^{12}\,{\rm cm}
              \simeq 43\,R_\odot,
\end{equation}
 where we adopted $M_{\rm B3V} = 7\,M_\odot$ and $V_{\rm wind} =
250$\,\kms.
An accretion rate can then be estimated from
\begin{equation}
\label{macc}
 \dot{M}_{\rm acc}
     \simeq \frac{\pi\,r_{\rm acc}^2}{(4\,\pi\,A^2)}\,\dot{M}_{\rm wind}
     \simeq 1 \times 10^{19}\,\xi^{1/2}\,{\rm g\,s}^{-1}, 
\end{equation}
where we used the values of $\dot{M}_{\rm wind}$ and $A$ estimated in
Sect.~\ref{feII_disc}.
This implies an accretion luminosity,
\begin{equation} 
L_{\rm acc} = \frac{G\,M_{\rm B3V}\,\dot{M}_{\rm acc}}{R_{\rm B3V}} 
    \simeq 9\,\xi^{1/2}\,L_\odot,
\end{equation}
where $R_{\rm B3V} = 4.5\,R_\odot$ has been assumed. 
This luminosity is small compared to the
intrinsic luminosity of a standard B3V star 
($\sim 1.9 \times 10^3\,L_\odot$, \citealp{sk82}). Thus, no
significant effect is expected to be observed in the global energetics of
the B3V companion, e.g., in the visual brightness.
However, after being dissipated via an accretion shock\footnote{The specific angular momentum
accreted with the matter is negligible in this case and
there is no chance to create an accretion disc arround the B3V
companion, contrary to the suggestion of \cite{goran08}.}
and thus producing a hot layer
above the star surface, the accretion luminosity is expected to be emitted  
mainly in the far-UV and/or soft X-rays. This can affect the
ionization of the expanding envelope of V838~Mon, 
i.e., produce significant ionization
of hydrogen. The     
accretion from the expanding matter is expected to become effective just  
before and during the November/December~2006 eclipse.
We suggest that this is the reason for the rapid increase in
emission of the Balmer lines, particularly in H$\alpha$, observed in the
second half of 2006. We note
that the luminosity in the Balmer emission lines, dominated by H$\alpha$, 
as observed by \cite{mun07} during the eclipse, was
$\sim 3.5\,L_\odot$ (assuming $E_{B-V} = 0.9$ and a 8~kpc distance to
V838~Mon). Thus, the accretion, as estimated above, can provide enough energy
to explain the observed luminosity in the Balmer lines.

\begin{acknowledgements} 
The research reported on in this paper has been
supported by the Polish Ministry of Science and Higher Education under
grant no. N203 004 32/0448, for which the authors are grateful.

\end{acknowledgements}

\end{document}